\documentclass[12pt]{iopart}
\usepackage{graphicx}
\usepackage{bm}
\usepackage{mathrsfs}
\usepackage{color}
\usepackage[dvipdfm]{hyperref}

\begin{document}

\title[]{Pseudospin dynamics in multimode polaritonic Josephson junctions}

\author{$^{1}$G. Pavlovic,$^2$G. Malpuech,$^{3,4}$I.A. Shelykh}

\address{$^1$International Institute of Physics, Av. Odilon Gomes de Lima, 1722, CEP 59078-400 Capim Macio Natal, RN, Brazil }

\address{$^2$LASMEA, UMR CNRS-Universit\'e Blaise Pascal 6602, 24
Avenue des Landais, 63177 Aubi\`ere Cedex France}

\address{$^3$Physics Department, University of Iceland, Dunhaga-3,
IS-107, Reykjavik, Iceland}

\address{$^3$ A.F. Ioffe Physico-Technical Institute of RAS, 194021 St.
Petesburg, Russia}

\ead{shelykh@raunvis.hi.is}
\begin{abstract}
We analyzed multimode Josephson junctions with exciton-polaritons (polaritonic Josephson junctions) when several coupling mechanisms of fundamental and excited states are present. The applied method is based on Keldysh-Green function formalism and takes into account polariton pseudospin. We found that mean value of circular polarization degree in intrinsic Josephson oscillations and microscopic quantum self-trapping follow an oscillator behavior whose  renormalizes due to inter-mode interactions. The effect of an additional  transfer of particles over junction barrier occurring in multimode approximation in combination with common Josephson tunneling is discussed in regime of dynamical separation of two polarizations.
\end{abstract}


\section{Introduction}

After theoretical prediction \cite{JosephsonBD} and experimental detection \cite{Likharev} of Josephson tunneling between two superconductors separated by a thick insulator under application of an external voltage, analogous quantum oscillations were found also in a superfluid system: two vessels of superfluid helium connected by a nanoscale aperture \cite{Pereverzev}.
Similar type of dynamics was observed for two weakly coupled atomic Bose-Einstein condensates (BECs) created in a double-trap potential. Differently from superconductors and superfluid helium in Josephson effect for atomic BECs inter-particle interactions play essential role \cite{Albiez}. An an-harmonic behavior occurs in these systems additionally to common Josephson oscillations. Under certain initial conditions BECs  localize in one of the traps formed by external potential due to suppression of the tunneling current \cite{Smerzi}. The phenomenon is known as macroscopic quantum self-trapping (MQST). It is a consequence of domination of nonlinearities, e.g. interactions, over Josephson coupling in the system. In some other range of the initial conditions delocalized phase is formed with unsuppressed oscillations between the traps.

In general, such a system - interacting BECs in a Josephson junction (JJ) is described by well-known Hamiltonian
\begin{equation}
H = {H_0}(z(0),\theta (0)) = \Lambda \frac{z(t)^2}{2} - \cos \theta(t)\sqrt {1 - {z(t)^2}} \label{Hamiltonian1}
\end{equation}
written in terms of  population imbalance $z(t)=(N_{1}(t)-N_{1}(t))/(N_{1}(t)+N_{1}(t))$, where $N_{1}$ and $N_{2}$ are populations in trap one and trap two, and  phase difference $\theta (t)$ between  BECs. Two-mode approximation including only lowest energy states of symmetric double-trap is used in Hamiltonian (\ref{Hamiltonian1}) \cite{Smerzi} and contains single parameter $\Lambda=U_{0}N_{T}/2J$. In the last expression $J$ figures as Josephson coupling constant which is equal to the  difference in energies of symmetric and anti-symmetric states of the double-trap potential is less then zero, as the symmetric state is usually lower in energy then the anti-symmetric one. $N_{T}$ is total population of particles in both traps which interact with energy $U_{0}$.

\begin{figure}
\begin{center}
\includegraphics[width=0.70\linewidth]{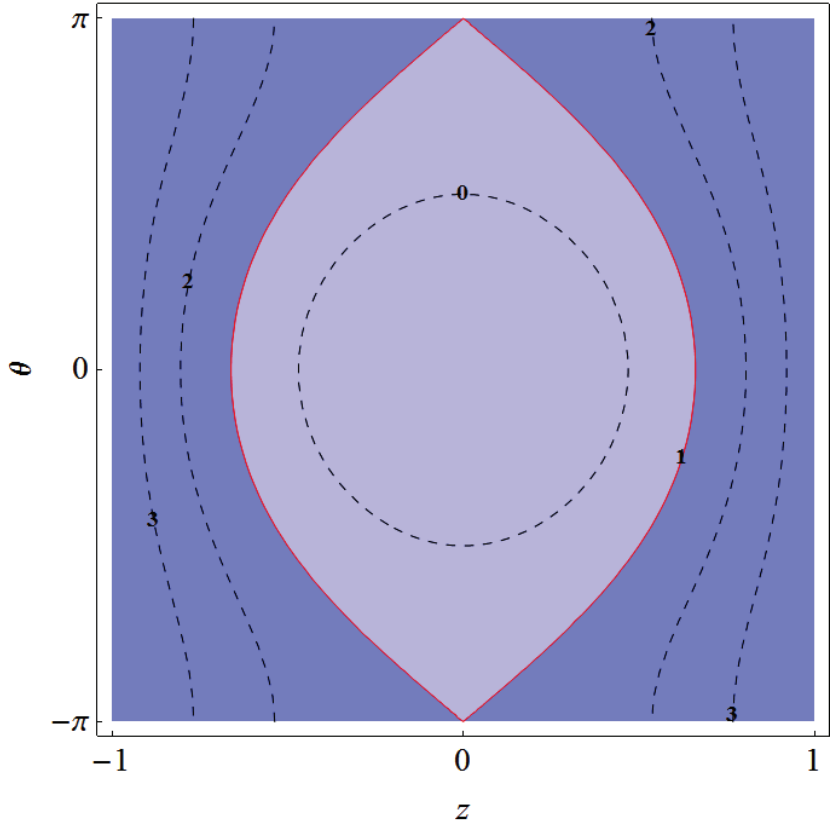}
\caption{ \label{Contour} Hamiltonian (\ref{Hamiltonian1}) in the phase space (z,$\theta$). Contour lines for $H$=0,1,2 and 3 are labeled. Separatrix is shown with full/red line}
\end{center}
\end{figure}

Inspecting the phase-space diagram ($z$,$\theta$) of Hamiltonian (\ref{Hamiltonian1}) one can observe two distinct regions separated by separatrix line $H=1$ ( full/red contour in Fig. \ref{Contour}). Fixing the value of $H$ in (\ref{Hamiltonian1}) by choosing initial conditions $z(0)$ and $\theta (0)$, delocalized regime will establish if $H<1$. It is characterized by the population imbalance being zero in average ($<z(t)>=0 $) for the system's cyclic motion on closed orbits (see Fig. \ref{Contour}). Approaching the separatrix from the inner side the orbits starts to deviate from regular circles as nonlinearities start to be important. Outside the separatrix where $H>1$ the population imbalance evolves along open lines with small oscillatins around constant mean value $<z(t)>=const$ indicating the transition to MQST effect.

Besides cold atoms there is an another kind of particles which can undergo a transition into collective state with properties similar to Bose- Einstein condensate (BEC). These are exciton-polaritons whose condensation has been repported several years ago in in CdTe microcavities \cite{KasprzakNature,Balili} at about 20 K and later on in GaN-based  microcavities even at room-temperatures \cite{Levrat,Christopoulos}. The possibility to obtain BEC at such conditions is extremely interesting from fundamantal point of view. Besides, it opens a way for design of novel optoelectronic components, such as polariton lasers \cite{Imamoglou} and various devices based on polariton superfluidity \cite{LiewPE}

Polaritons are two-dimensional quasi-particles appearing due to strong coupling of excitons in semiconductor quantum-well(s) and photons confined within a microcavity structure. Peculiar properties of exciton-polaritons  are consequences of their hybrid, half-light half-mater nature coming from photonic  and excitonic components respectively. In particular, ultra-small effective mass of polaritons, which is typically four or five orders of magnitude less then the free electron mass, makes quantum collective phenomena very pronounced in polaritonic systems and leads extremely high critical temperatures of the polariton BEC transition.

For the purposes of our further discussion, another property of polaritons is very important. That is projection of total angular momentum $J_z^{pol}$ of a polariton state on the structure's growth axis (chosen as $z$ direction) \cite{ShelykhReview}. It can take two values: $J_z^{pol}= \pm 1$ corresponding to mixing of bright excitons ($J_z^{exc}= \pm 1$) and photons with right or left circular polarizations. Dark excitons with $J_z^{exc}= \pm 2$ are uncoupled from cavity mode due to optical selection rules and thus do not contribute to the formation of a polariton doublet. From the point of view os spin polariton is thus analogical to electron, as both of them are two-level systems. Consequently one can apply pseudospin formalism for description of spin dynamics of the polaritons. It is convenient to represent a pseudo spin state of polariton system by a point on Poincar\'e sphere (also known as Bloch sphere). Importantly, polariton pseudospin unambiguously defines the polarization state of photoemission from the cavity. The states lying at the poles of Poincar\'e sphere will correspond to circular polarization of the polaritons, the states on equator- to linear polarizations and all other states to elliptical polarizations.

Being bosons, exciton-polaritons should manifest phenomena related to Josephson effect. Polaritonic Josephson junctions (PJJ) were considered theoretically in Refs. \cite{Wouters,Wouters1,Savona} and experimentally in  Ref.\cite{Lagoudakis}. In these articles polariton pseudospin was neglected, and and Josephson coupling occurred between the polaritons located in spatially separated traps. The introduction of polarization degree of freedom makes Josephson dynamics far more rich. Indeed, in addition to the coupling between states localized in different traps, two polarization states in single asymmetric trap can be coupled as well by mechanism analogical to TE- TM splitting \cite{Aleiner}. Josephson-type oscillations can occur between different polarizations, the phenomenon which was called intristic Josephson effect \cite{ShelykhClermont}. The interplay between intrinsic and extrinsic Josephson dynamics can lead to dynamic separation of different polarizations in the real space and other intriguing phenomena \cite{ShelykhClermont}.

The role of the polarization coupling in PJJ was studied for several configurations \cite{Dima,Dean,Einar}. In most of them so called two-mode approximation was applied, in which only two lowest levels, one in each of two traps, were considered.  An exception is the reference \cite{Einar} where several modes were included in analysis in order damping of Josephson oscillations due to the interactions of the polaritons with acoustic phonons. However, even in this case the separation between two fundamental modes and excited ones was considered to be much greater then characteristic values of blueshifts provided by polariton- polariton interactions, and transitions to the excited modes were only possible due to absorbtion of phonons.

In this paper we analyze multimode PJJ with confining potential created in such a way that interactions between the lowest and excited levels are not negligible. This kind of coupling have been already considered for Josephson Junctions based on atomic BECs \cite{Martinez}.  Our goal is to extend the analysis of Ref.\cite{Martinez} for the case of cavity polaritons accounting the polarization degree of freedom, and clarify the effects of the coupling between fundamental and excited modes on nonlinear polarization dynamics of PJJs.

\section{Model}

In Ref.\cite{Martinez} Keldysh-Green functions technique was employed to study dynamics for atomic Josephson Junction with multimode structure. Here we shall  follow similar approach for description of PJJs accounting for the spin degree of freedom. Because of spinor nature of polariton condensate in addition to normal Josephson coupling we have to consider also and the coupling of two polarization components, normally irrelevant for atomic JJs.

We start by expanding exciton-polariton field operators $\hat \Psi _{\sigma^{\pm}}(r,t)$ over complete set of eigenstates ${\{f_1,f_2,...\}}$ of a double-well potential
\begin{eqnarray}
\hat \Psi _{ \sigma^{\pm}}(r,t) = f_{1}(r)\hat a_{1 \sigma^{\pm}}(t)+f_{2}(r)\hat a_{2 \sigma^{\pm}}(t)+\sum\limits_{n = 3}^\infty  {{f_n}(r)\hat b _{n \sigma^{\pm}}(t)}  \label{doublewellexpansion}
\end{eqnarray}
where $\hat a_{1 \sigma^{\pm}}$ and $\hat a_{2 \sigma^{\pm}}$ denote annihilation operators for two lowest modes of PJJ localized in the traps one and two, and $b_{n \sigma^{\pm}}$ stands for the anihilation operator on  excited level $n$ taking values. $\sigma^\pm$ denotes circular polarization of the state.

The system we study is described by the Hamiltonian which can be represented as a sum of three terms:
\begin{equation}
\hat H =  \hat H_{{\sigma ^ + }} +  \hat H_{{\sigma ^ - }} +  \hat H_{{\sigma ^ + }{\sigma ^ - }}  \label{fullHamiltonian}.
\end{equation}

The terms conserving z- projection of pseudospin ${ \sigma_+ }$ and ${\sigma_-  }$ denoted by $\hat H_{{\sigma ^ + }}$ and $\hat H_{{\sigma ^ - }}$ are given by following expressions
\begin{eqnarray}
\fl \hat H_{{\sigma ^ {\pm}}}=
\int {{d^2}r \hat \Psi _{{\sigma ^ {\pm}}}^ \dag (r,t)} ( { - \frac{{{\hbar ^2}}}{{2m}}\Delta  + {V_{ext}}( r,t)} ){ \hat \Psi _{{\sigma ^ {\pm}}}}(r,t) \label{SQHamiltonian} \\
+  \frac{1}{2}\int {{d^2}} r\int {{d^2}} {r^{'}}\hat \Psi _{{\sigma ^ {\pm}}}^ + (r,t)\hat \Psi _{{\sigma ^ {\pm}}}^\dag (r^{'},t)V(r - r^{'}){ \hat \Psi _{{\sigma ^{\pm}}}}(r^{'},t){ \hat \Psi _{{\sigma ^ {\pm}}}}(r,t). \nonumber
\end{eqnarray}
The effective mass of polaroton is denoted by $m$ and $V_{ext}(r,t)$ stands for external double-well potential. $V(r,r^{'})=g \delta (r-r^{'})$ is contact interaction described by delta function and interaction constant  can be estimated as $g\approx E_Ba_B^2$ with $E_B$ and $a_B$ being being exciton binding energy and Bohr radius of the exciton respectively \cite{CiutiMatrEl}.

The last term in the Eq. (\ref{fullHamiltonian}) accounts for  the Josephson- type coupling of the particles having opposite pseudospin projections. It can be viewed as a consequence of one presence of  effective in-plane magnetic field $\Omega(r)$ arising from the asymmetry of the structure \cite{Aleiner} and acting on polaritons spins and can be represented as
\begin{equation}
\hat H_{{\sigma ^ + }{\sigma ^ - }} = \int {{d^2}r \hat \Psi _{{\sigma ^ {\pm}}}^\dag (r,t)} \Omega(r){ \hat \Psi _{{\sigma ^ {\mp}}}}(r,t).
\label{magHamiltonian}
\end{equation}
After subtitution of the expensions (\ref{doublewellexpansion}) into the Hamiltonians (\ref{SQHamiltonian}) and (\ref{magHamiltonian}) the starting, i.e. full Hamiltonian (\ref{fullHamiltonian}) can be recast as
\begin{equation}
\hat H =  \hat H_{}^{0} +  \hat H_{}^{exc} +  \hat H_{}^{int}.
\label{grouppedH}
\end{equation}
where the first term describes the dynamics of the four  fundamental modes (accounting to polarization degreee of freedom), the second term describes the dynamics of the delocalized excited modes and the last term corresponds to the coupling between fundamantal and excited modes. These terms read:
\begin{eqnarray}
\fl  \hat H_{}^{0} = {E_{0}}\sum\limits_{i;\sigma }^{} {\hat a _{i\sigma }^\dag {\hat a _{i\sigma }}}  + J_0\sum\limits_\sigma ^{} ( \hat a_{1\sigma }^\dag {\hat a _{2\sigma }} + \hat a _{2\sigma }^\dag {\hat a _{1\sigma }}) + \frac{{{U_{0}}}}{2}\sum\limits_{i;\sigma }^{} {\hat a_{i\sigma }^\dag \hat a_{i\sigma }^\dag {\hat a_{i\sigma }}} {\hat a_{i\sigma }} +
{\Omega _{0}}\sum\limits_{i;\sigma }^{} {\hat a_{i\sigma }^\dag {\hat a_{i - \sigma }}}; \label{HamiltonianBEC} \\
\fl \hat H_{}^{exc} =  \sum\limits_{n,m;\sigma }^{} {({E_n} + {U_{nm}}}  < \hat b_{m\sigma }^\dag {\hat b_{m\sigma }} > )\hat b_{n\sigma }^\dag {\hat b_{n\sigma }}+{\Omega _n}\sum\limits_{n;\sigma }^{} {\hat b_{n\sigma }^\dag {\hat b_{n - \sigma }}} \label{HamiltonianEXC} \\
+ \frac{{{U_{nm}}}}{2}\sum\limits_{n,m;\sigma }^{} {\left( { <\hat b_{m\sigma }^\dag \hat b_{m\sigma }^\dag  > {\hat b_{n\sigma }}{\hat b_{n\sigma }} +  < {\hat b _{m\sigma }}{\hat b_{m\sigma }} > \hat b_{n\sigma }^\dag \hat b_{n\sigma }^\dag } \right)}   ; \nonumber \\
\fl \hat H_{}^{int} =  K_n\sum\limits_{i,n;\sigma }^{} {[\frac{1}{2}\left( {\hat a_{i\sigma }^\dag \hat a_{i\sigma }^\dag {\hat b_{n\sigma }}{\hat b_{n\sigma }} + h.c.} \right) + 2\hat a_{i\sigma }^\dag \hat a_{i\sigma }^{}\hat b_{n\sigma }^\dag {\hat b_{n\sigma }}} ] + \label{hamiltonianint} \\
+ J_n\sum\limits_{n;\sigma }^{} {[2(\hat a_{1\sigma }^\dag \hat a_{2\sigma }^{} + \hat a_{2\sigma }^\dag \hat a_{1\sigma }^{})\hat b_{n\sigma }^\dag {\hat b_{n\sigma }} + (\hat a_{1\sigma }^\dag \hat a_{2\sigma }^\dag {\hat b_{n\sigma }}{\hat b_{n\sigma }} + h.c.)} ]. \nonumber \\ \nonumber
\end{eqnarray}

In the above expressions $E_n$ are energies of the modes, $J_0$ is a Josephson coupling strength between two fundamental modes localized in right and left traps of the double- well potential, $\Omega_n$ are coupling strengthes between states of different polarizations at level $n$ corresponding to intrinsic Josephson effect, $U_0$ describes polariton- polariton interaction in the fundamental modes, $U_{mn}$- interactions in excited modes, $K_n$- interactions between fundamental and excited modes and $J_n$ describes the renormalization of the Josephson tunneling due to the interaction between fundamental and excited modes. The parameters entering into Hamiltonian \ref{grouppedH} are related to those entering into the initial Hamiltonian \ref{SQHamiltonian} as

\begin{eqnarray}
{E_{0(n)}} = \int {{d^3}} rf_{1,2(n)}^*(r)\left( { - \frac{{{\hbar ^2}}}{{2m}}\Delta  + V_{ext}(r)} \right){f_{1,2(n)}}(r), \label{KE}
\end{eqnarray}
\begin{eqnarray}
{J_0} = \int {{d^3}} rf_{1,2}^*(r)\left( { - \frac{{{\hbar ^2}}}{{2m}}\Delta  + {V_{ext}}(r)} \right){f_{2,1}}(r), \label{JC}
\end{eqnarray}
\begin{eqnarray}
{U_{0}} = \int {{d^3}r{d^3}{r^{'}}f_{1,2}^*(r)f_{1,2}^*(r^{'})V(r,r^{'}){f_{1,2}}(r^{'}){f_{1,2)}}(r)} = \nonumber \\
= g\int {{d^3}} r{\left| {{f_{1(2)}}(r)} \right|^4}, \label{II}
\end{eqnarray}
\begin{eqnarray}
{U_{nm}} = g\int {{d^3}} r{{f_{m}^{*2}(r)f_{n}^{2}}(r)} \label{III},
\end{eqnarray}
\begin{eqnarray}
{\Omega_{0(n)}} = g\int {{d^3}} r{{f_{1\sigma,2\sigma(n\sigma)}^{*}(r)\Omega(r)f_{1-\sigma,2-\sigma(n-\sigma)}(r)}} \label{VI},
\end{eqnarray}
\begin{eqnarray}
{K_{n}} = g\int {{d^3}} r{{f_{1,2}^{*2}(r)f_{n}^{2}}(r)}=g\int {{d^3}} r{\left| {{f_{1,2}(r)f_{n}}(r)} \right|^2} \label{IV},
\end{eqnarray}
\begin{eqnarray}
{J_{n}} = g\int {{d^3}} r{{f_{1}^{*}(r)f_{2}}(r)\left| {{f_{n}}(r)} \right|^2}=g\int {{d^3}} r{f_{1}^{*}(r)f_{2}^{*}(r)f_{n}^{2}(r)} \label{V}.
\end{eqnarray}

In the part corresponding to the excited states $\hat H_{exc}$ the interactions are treated using the mean field approximation \cite{Griffin}
\begin{equation}
\fl \hat b _{m\sigma }^\dag \hat b _{n\sigma }^\dag {\hat b _{n\sigma }}{\hat b _{m\sigma }} = 2 < \hat b _{m\sigma }^\dag {\hat b _{m\sigma }} > \hat b _{n\sigma }^\dag {\hat b _{n\sigma }}
+ <\hat b _{m\sigma }^\dag \hat b _{m\sigma }^\dag  > {\hat b _{n\sigma }}{\hat b _{n\sigma }} +  < {\hat b _{m\sigma }}{\hat b _{m\sigma }} > \hat b _{n\sigma }^\dag \hat b _{n\sigma }^\dag .  \label{approximation}
\end{equation}

In order to obtain a closed system of dynamical equations describing interacting PJJ we write Keldysh propagators of the system in the following representation
\begin{equation}
{{\tilde G}_{\alpha \beta}}(t,{t^{'}}) = \left( {\begin{array}{*{20}{c}}
   {{{\hat G}_{{\alpha \beta}\sigma \sigma }}(t,{t^{'}})} & {{{\hat G}_{{\alpha \beta}\sigma  - \sigma }}(t,{t^{'}})}  \\
   {{{\hat G}_{{\alpha \beta} - \sigma \sigma }}(t,{t^{'}})} & {{{\hat G}_{{\alpha \beta} - \sigma  - \sigma }}(t,{t^{'}})}  \\
\end{array}} \right), \label{fullmatrix}
\end{equation}
where the general indices $\alpha$ and $\beta$ become $i$ or $j$ for the ground states and take values $1$ or $2$ as there are two lowest lying modes. The excited states are counted by associating  $m$ or $n$ to both $\alpha$ and $\beta$. As previously indices $\pm \sigma $ denotes pseudospin degrees of freedom. The elements of the above matrix (\ref{fullmatrix})  are themselves $2 \times  2$ block-matrices of the form
\begin{equation}
{i\hat G_{\alpha\beta\sigma \sigma }}(t,{t^{'}}) =  \left( {\begin{array}{*{20}{c}}
   { G_{\alpha\beta\sigma \sigma }(t,t^{'})} & { F_{\alpha\beta\sigma \sigma }(t,t^{'}) }  \\
   { \bar{F}_{\alpha\beta\sigma \sigma }(t,t^{'}) } & { \bar{G}_{\alpha\beta\sigma\sigma }(t,t^{'})}  \\
\end{array}} \right). \label{blockmatrix0}
\end{equation}
with the elements being, for example for the reservoir modes
\begin{eqnarray}
G_{\alpha\beta\sigma \sigma} (t,t^{'})= < {\mathscr{T}}{\hat b_{l\sigma }}(t)\hat b_{m\sigma }^\dag ({t^{'}}) > \\ \label{blockmatrix1}
F_{\alpha\beta\sigma \sigma} (t,t^{'})= < {\mathscr{T}}{\hat b_{l\sigma }}(t)\hat b _{m\sigma }^{}({t^{'}}) >. \\ \label{blockmatrix2} \nonumber
\end{eqnarray}
and similarly for the fundamental modes localized on the traps 1 and 2.
Time-ordering $\mathscr{T}$ in the previous formulae is performed on the Keldysh contour \cite{Keldysh} and appears because of non-adiabatic switching of Josephson coupling in the initial time instant. As it is not possible to guarantee the behavior of the system with such a kind of the irreversibility for asymptotically large times \cite{Haug} the time contour  in Keldysh formulation is adapted so that system evolves forwardly on time axis to some time and then from this point it makes backward evolution to the initial state. Any Keldysh time-ordered product of two operators, for example, operators $\hat b_{n}$ and $\hat b_{m}$, has form
\begin{equation}
< {{\mathscr{T}}[ {{\hat b _m}(t){\hat b _n}(t^{'})} ]} >  = \theta (t,t^{'})< {{\hat b _m}(t){\hat b _n}(t^{'})} >
+ \theta (t^{'},t)< {{\hat b _n}(t^{'}){\hat b _m}(t)} > .  \label{lg}
\end{equation}
where $\theta (t,t^{'})$ is "step" function defined on the Keldysh contour for two arbitrary times $t$ and $t^{'}$ in way that is one always when the first argument is later then the second one, and zero otherwise. In this sense the first Green function in the formula (\ref{lg}) is called "greater", denoted usually with $F^{>}$ and the second one  is "lesser" Green function  $F^{<}$. They act only on the forward or the backward Keldysh contour branch, respectively.

Using Wigner transformation, the matrix elements $G_{\alpha \beta \sigma \sigma }(t,t^{'})$ can be written in terms of center of "mass" and "relative" time coordinates $T=(t+t{'})/2$ and $\tau=t-t{'}$. We are interested here in external and internal Josephson dynamics which are much slower processes then the others occuring in the system so that we can work in the limit in which $\tau = 0$. The new matrix elements $G_{\alpha \beta \sigma \sigma }(T,\tau)$ will then only depend on the macroscopical time $T$ \cite{Martinez}. According to the expression (\ref{lg}) we will then deal only with "lesser" functions $F^{<}(T)$ and $G^{<}(T)$.

Equations of motion techniques combined with use of the mean- field approximation allows us to obtain the closed system of equations for Green functions defined by Eq.\ref{blockmatrix0}:

\begin{eqnarray}
\fl i\frac{{dG_{n\sigma n\sigma }^ < }}{{dT}} = {\Xi _{n\sigma }}\bar F_{n\sigma n\sigma }^ <  - {\bar \Xi _{n\sigma }}F_{n\sigma n\sigma }^ <  - {\Omega _n}(G_{n\sigma n - \sigma }^ <  - G_{n - \sigma n\sigma }^ < ); \label{1} \\
\fl (i\frac{d}{{dT}} + {\Upsilon _{n\sigma }}-{\Upsilon _{n - \sigma }})G_{n\sigma n - \sigma }^ <  =  2{\Xi _{n\sigma }}\bar F_{n\sigma n - \sigma }^ <- 2{\bar \Xi _{n\sigma }}F_{n\sigma n - \sigma }^ <  - {\Omega _n}(G_{n\sigma j\sigma }^ <  - G_{n - \sigma n - \sigma }^ < ); \label{2} \\
\fl ( {i\frac{d}{{dT}} - 2{E_{n\sigma }} - 2{\Upsilon _{n\sigma }}} )F_{n\sigma n\sigma }^ <  = 2{{\bar \Xi }_{n\sigma }}G_{n\sigma n\sigma }^ <  + 2{\Xi _{n\sigma }}\bar G_{n\sigma n\sigma }^ <  - {\Omega _n}F_{n\sigma n - \sigma }^ < ; \label{3} \\ \nonumber
\fl (i\frac{d}{{dT}} + {2E_{n\sigma }} + {\Upsilon _{n\sigma }}+ {\Upsilon _{n -\sigma }})F_{n\sigma n - \sigma }^ <  = -2({\Xi _{n\sigma }-\Xi _{n -\sigma }})G_{n\sigma n - \sigma }^ <  - {\Omega _n}(F_{n\sigma n\sigma }^ <  + F_{n - \sigma n - \sigma }^ < ). \label{4} \\
\end{eqnarray}
where $n=3,4,...$ stands for the $n$ th excited level. Dynamics of the fundamental states $i=1,2$ is given by following expressions:
\begin{eqnarray}
\fl i\frac{{dG_{i\sigma i\sigma }^ < }}{{dT}} = ( {{E_0} + {U_0}G_{i\sigma i\sigma }^ <  + 2K_n\sum\limits_{n\sigma } {{G_{n\sigma n\sigma }^ <}} } )G_{i\sigma i\sigma }^ <  + (J + 2{J_{n}}\sum\limits_{n\sigma } {{G_{n\sigma n\sigma }^ <}} )G_{i\sigma j\sigma }^ <  +\label{5}  \\ \nonumber
+ i( {K_n\bar F_{i\sigma i\sigma }^ <  + {J_n}\bar F_{j\sigma j\sigma }^ < } )\sum\limits_{n\sigma} {\bar F_{n\sigma n\sigma }^ < }  - {\Omega _0}G_{i\sigma i - \sigma }^ <.
\end{eqnarray}
with $i,j=1,i\neq j$.

Self-energies $\Upsilon $ and $\Xi$ read
\begin{eqnarray}
\fl {-i\Upsilon _{n\sigma }} = U^{*}\sum\limits_m {G_{m\sigma m\sigma }^ < }
+2K_n({G_{1\sigma }^ <} + {G_{2\sigma }}^ <) + 2{J_{n}}(G_{1\sigma 2\sigma }^ <  + G_{2\sigma 1\sigma }^ <),  \\ \label{SFY}
\fl -i{\Xi _{n\sigma }} = \frac{{{U^{*}}}}{2}\sum\limits_m {F_{m\sigma m\sigma }^ < }  + \frac{K_n}{2}(F_{1\sigma 1\sigma }^ < + F_{2\sigma 2\sigma }^ <) + {J_{n}}F_{1\sigma 2\sigma }^ <.\label{SFO} \\ \nonumber
\end{eqnarray}
The normal and anomalous self-energies represent the energy renormalizations entering to the diagonal and off-diagonal propagators in the matrix (\ref{blockmatrix0}) due to particle- particle interactions present in PJJ. For simplicity we take them diagonal in excited level index $n$ neglecting collisions between particles situated at different reservoir levels, $U_{nm}=U^{*}\delta_{nm}$.

The system of (\ref{1})-(\ref{SFY}) is the closed system of nonlinear first order ordinary differential equations which can be solved numerically. The corresponding analysis is presented in the next section.

\section{Results and Discussions}

We consider a PJJ inside GaAs- based quantum microcavity with parameters similar to those used in Ref.\cite{ShelykhClermont}. We numerically studied the system of equations (\ref{1})-(\ref{SFY}) in order to analyze  the effects of the interactions between fundamental and excited states given by parameters $K_n$ and $J_{n}$ on various types of Josephson deynamics.  In the reference \cite{Martinez} it was found that for atomic condensates such kinds of interactions in general lead to chaotization of the Josephson oscillations after some initial period of regular dynamics. The effect is due to the intensive exchange of the particles between fundamental states and multi- mode reservoir of the excited states. Here we consider time intervals smaller then those necessary for the transition to chaotic regime. The reason is that polaritons have finite lifetimes, and in the regime of pulsed excitation they will simply disappear before the system will demonstrate the characteristics of chaos \cite{ShelykhClermont}.

\begin{figure}
\begin{center}
\includegraphics[width=0.99\linewidth]{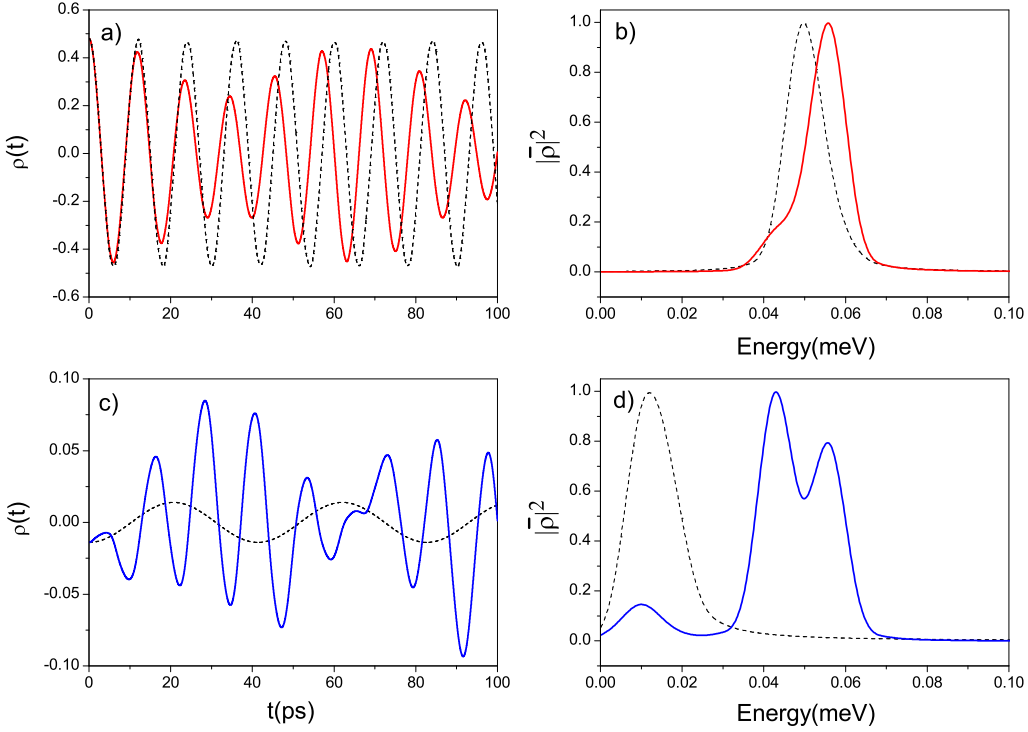}
\caption{ \label{figure2} Panels a) and b) Profile of the intrinsic Josephson oscillations of circular polarization degree at fundamental states and its Fourier power spectrum. Panels c) and d) Profile of the intrinsic Josephson oscillations of circular polarization degree at first excited state and its Fourier power spectrum. Dashed lines correspond to the absence of the coupling between fundamantal modes and excited states, $K_n=0$. Solid lines correspond to the case $K_n=0.3U_0$.}
\end{center}
\end{figure}

The quantity
\begin{equation}
\rho_{n}(T)=\frac{G_{n\sigma^+ i\sigma^+ }^ <(T)- G_{n -\sigma^- i-\sigma^-}^ <(T)}{G_{n\sigma^+ i\sigma^+ }^ <(T)- G_{n -\sigma^- i-\sigma^-}^<(T) }
\label{degree}
\end{equation}
describes circular polarization degree at the state n. Its dynamics is shown at Figure \ref{figure2}. For a while, we neglected the extrinsic Josephson coupling puting $J_0=0,J_n=0$. Panels a) and b) show a profile of the oscillations of circular polarization degree in the fundamental states corresponding to intrinsic Josephson effect and corresponding Fourier spectrum. The dashed line correspond to the case when coupling to excited states is switched off, while solid line accounts for this coupling. One sees, that introduction of the term $K_n$ slightly renormalizes the frequency of the oscillations. However, the effect remains quite weak, as population of the excited levels is sufficiently small, less then ten percents of total number of particles. On the contrary, the effect of coupling becomes more pronounced if one monitors the intrinsic Josephson dynamics on the excited states, shown at panels  c) and d). The interaction with fundamental modes changes the oscillation pattern on excited modes quite radically. The effect is clearly seen at panel d) showing Fourier power spectrum of the oscillations. Account for the terms $K_n$ leads to the appearance of higher harmonics in the spectrum. Besides, in place of the intrinsic Josephson oscillations with zero time average,  novel regime establishes in which  $<\rho(T)> \neq 0$. In the absence of coupling $K_n$ the frequencies of the intrinsic oscillations in the ground state and the reservoir are different (dashed lines in panels a) and c)) although polarization couplings on these levels are equal $\Omega_0=\Omega_n$. It is result of the renormalization of the oscillation frequency for the fundamental mode due to nonlinearities, which is negligible in the low populated reservoir. Account of the reservoir- fundamental mode coupling makes the frequencies of intrinsic Josephson oscillations comparable for all modes of the system.

\begin{figure}
\begin{center}
\includegraphics[width=0.99\linewidth]{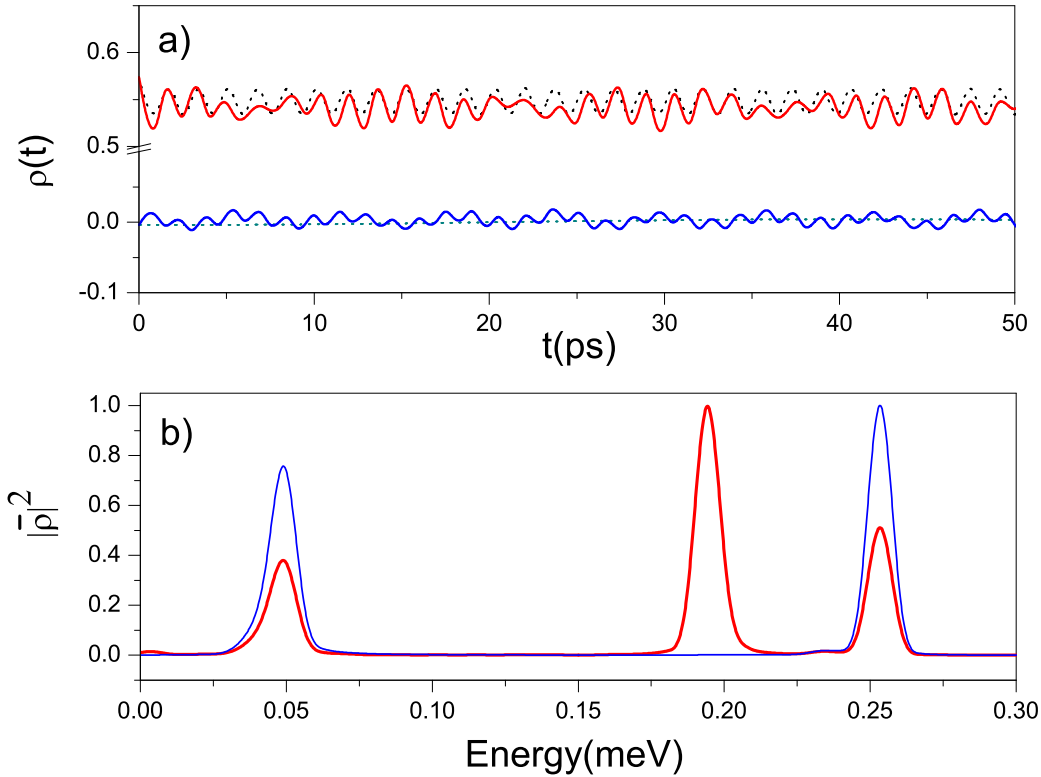}
\caption{ \label{figure3} Top of the panel a) shows the polarization degree of the intrinsic Josephson oscillations in the fundamental mode. Bottom of the panel a) shows the same quantity in the reservoir. Dashed lines: $K_n = 0$. Thick/blue and solid/red lines stands for $K_3= 0.3U_0$. Panel b) shows corresponding Fourier power spectrum when the interactions are present (red for the fundamental mode, blue for the reservoir).}
\end{center}
\end{figure}

Fig. \ref{figure3} illustrates behavior of the polarization degree when ground state MQST phase in the intrinsic Josephson effect interacts with the reservoir. As compare the situation analyzed in Fig \ref{figure2} particle-particle interactions characterized by a parameter $U_0$ dominate over Josephson coupling $\Omega$ in the fundamental mode . The profiles of the oscillations of circular polarization degree for the fundamental modes are shown in the top of the panel a). The solid line corresponds to the case $K_n \neq 0$ and dashed one for $K_n= 0$.  The dashed line is characterized by a single peak in the Fourier spectrum, usual for MQST regime in two- mode PJJ. On the contrary, the introduction of the coupling with excited level leads to the appearance of the two additional peaks in Fourier spectrum as it is shown at the panel b).  The similar trends can be seen for oscillations in the excited level, illustrated by lower curves at panel a) and blue line at panel b).  

\begin{figure}
\begin{center}
\includegraphics[width=0.99\linewidth]{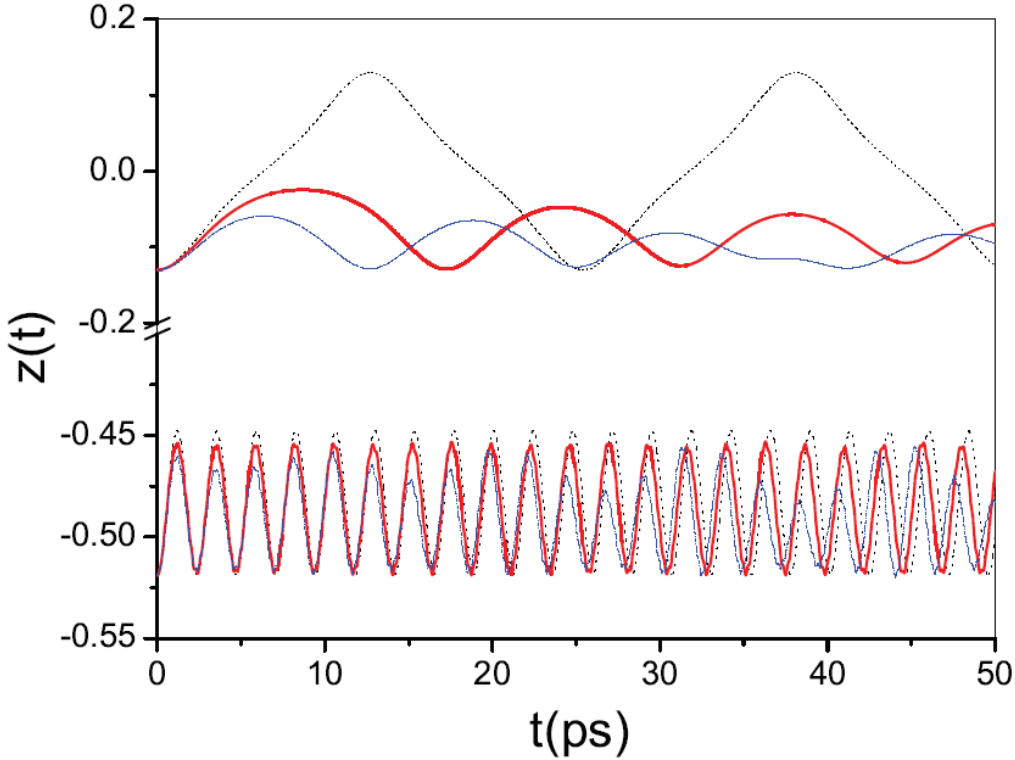}
\caption{ \label{figure4} Upper panel: Josephson oscillations of the  population imbalances between two traps for minor polarization component. Lower panel: Josephson oscillations of the  population imbalances between two traps for major polarization component. $J_0=50 \mu eV$. Dashed line: no reservoir assisted coupling, $J_n=0$. Red line: $J_3=0.1U_0$, blue line $J_3=0.2U_0$.}
\end{center}
\end{figure}

Finally we consider the scenario in which two fundamental modes localized in traps one and two, initially  elliptically polarized, are coupled with each other by means of extrinsic Josephson tunneling $J$. It was already pointed out that spatial separation of polarization occurs in this case \cite{ShelykhClermont} for sufficiently high polariton concentrations and circular polarization degrees. In this regime the dominant polarization component is trapped in one of the wells (dotted line in lower panel of Fig. \ref{figure4}), while another one undergoes Josephson oscillations  (dotted line in upper panel of Fig. \ref{figure4}). In the presence of the higher modes there is an extra pseudospin-conserving exchange of the particles through the excited levels between the traps described by reservoir- assisted coupling term $J_n$ in Eq.\ref{hamiltonianint}). The influence of this term is illustrated in Fig.\ref{figure4} by solid red and blue lines corresponding to $J_n=0.1U_0$ and $J_n=0.3U_0$ respectively. For freely oscillating component the presence of the reservoir-assisted term leads to self-trapping. This transition is provided by competition of two effects: common Josephson tunneling $J$ and reservoir-assisted coupling given by $J_n$. As their signs are opposite ($J<0$ and $J_n>0$), increase of $J_n$ decrease in fact absolute value of the total Josephson coupling. This results in overwhelming of free oscillations by nonlinearities and establishing of MQST in the minor polarization component. The smaller is the total Josephson coupling the trapping becomes stronger (compare blue and red line of the upper panel). The dominant component rests relatively robust to the influence of the reservoir-assisted coupling. Note, that in hypothetical case of positive $J$ the result will be opposite to those considered here: the reservoir- assisting terms will increase the absolute value of the effective tunneling constant, thus contributing to the distruction of MQST regime for major polarization component.

\section{Conclusions}

In conclusion, we analyzed polarization dynamics in multimode Polartonic Josephson Junctions. We have found that the coupling between fundamental and excited modes changes the patterns of intrinsic and extrinsic Josephson oscillations, leading to the appearance of higher harmonics and transitions between Josephson and MQST regimes.

\section{Acknowledgment}

We thank Dr.D.D. Solnyshkov and Prof. N. Gippius for stimulating discussions. The work was supported by FP7 IRSES project "SPINMET" and Rannis "Center of excellence in polaritonics". I.A.S. acknowledges the support from COST "POLATOM" project and thanks Mediterranean Institute of Fundamental Physics for hospitality.

\end{document}